\documentclass{sig-alternate-05-2015}

\usepackage{graphicx}
\usepackage{times}
\usepackage{helvet}
\usepackage{courier}
\usepackage{booktabs}
\usepackage[numbers,sort]{natbib}
\usepackage{epsfig}
\usepackage{amssymb}
\usepackage{amsmath}
\usepackage{amsfonts}
\usepackage{breqn}
\usepackage{multirow}
\usepackage{array}
\usepackage{subcaption}
\usepackage{mathtools}
\usepackage{caption}
\usepackage{url}
\usepackage{float}
\usepackage{pifont}
\usepackage{balance}
\conferenceinfo{RecSys}{'16 Boston, U.S.A.}

\newcommand{\eq}[1]{Eq.~\ref{#1}}
\newcommand{\xhdr}[1]{\vspace{1mm}\noindent{{\bf #1}}}
\newcommand{\md}{\emph{Vista}}
\newcommand{\modelnamelong}{VIsually, Socially, and Temporally-aware Artistic recommendation}
\newcommand{\website}{\emph{Behance}}

\DeclareMathOperator*{\argmax}{arg\,max}

\CopyrightYear{2016} 
\setcopyright{acmcopyright}
\conferenceinfo{RecSys '16,}{September 15-19, 2016, Boston, MA, USA}
\isbn{978-1-4503-4035-9/16/09}\acmPrice{\$15.00}
\doi{http://dx.doi.org/10.1145/2959100.2959152}
\begin{document}

\numberofauthors{4}
\author{
\alignauthor
Ruining He\\
       \affaddr{UC San Diego}\\
       \email{r4he@cs.ucsd.edu}
\alignauthor
Chen Fang\\
       \affaddr{Adobe Research}\\
       \email{cfang@adobe.com}
\alignauthor 
Zhaowen Wang\\
       \affaddr{Adobe Research}\\
       \email{zhawang@adobe.com}
\and
\alignauthor 
Julian McAuley\\
       \affaddr{UC San Diego} \\
       \email{jmcauley@cs.ucsd.edu}
}



\title{Vista: A Visually, Socially, and Temporally-aware Model \\ for Artistic Recommendation}
\maketitle

\begin{abstract}
Understanding users' interactions with highly subjective content---like artistic images---is challenging due to the complex semantics that guide our preferences. On the one hand one has to overcome `standard' recommender systems challenges, such as dealing with large, sparse, and long-tailed datasets. On the other, several new challenges present themselves, such as the need to model content in terms of its visual appearance, or even social dynamics, such as a preference toward a particular artist that is independent of the art they create.

In this paper we build large-scale recommender systems to model the dynamics of a vibrant digital art community, \website, 
consisting of 
tens of millions of
interactions (clicks and `appreciates') of users toward digital art. Methodologically, our main contributions are to model (a) rich content, especially in terms of its visual appearance; (b) temporal dynamics, in terms of how users prefer `visually consistent' content within and across sessions; and (c) social dynamics, in terms of how users exhibit preferences both towards certain art styles, as well as the artists themselves.
\end{abstract}


\section{Introduction}

The power of recommender systems lies in their ability to model the complex preferences that people exhibit toward items based on their past interactions and behavior. To extend their expressive power, various works have made use of features such as temporal dynamics \cite{timeSVD}, social influence \cite{ma2008sorec}, or the content of the items themselves \cite{upsdowns}. Yet modeling and recommendation still remains challenging in settings where these forces interact in subtle and semantically complex ways.

In this paper, we propose new models for the social art website \website,\footnote{\url{https://www.behance.net/}}
an online art community with millions of users and artistic images. We seek to model both implicit preferences, such as users navigating between items, as well as explicit preferences, such as users liking or `appreciating' each other's content. On \website{} users are both content creators and content evaluators, meaning that there is a need to model who appreciates what, as well as who appreciates whom.

There are several aspects that make modeling this data interesting and challenging. First is the need to model \emph{content}, 
in particular the visual appearance of items, which are both high-dimensional and semantically complex.
Second is the need to model temporal and social dynamics, in terms of users' tendency to interact with consistent content within and across sessions, and their preferences towards individual artists as well as art styles.
And third is simply the scale and sparsity of the data involved; with data on the order of millions of users and items, and tens of millions of interactions we must expend considerable effort developing methods that scale; this issue is exacerbated when modeling visual content, due to the high-dimensional representations of the items involved.

Methodologically, our work primarily builds upon two methods, \emph{FPMC} (`factorized personalized markov chains,' \cite{rendle2010fpmc}) and \emph{VBPR} (`visual bayesian personalized ranking,' \cite{VBPR}). FPMC models the notion of `smoothness' between subsequent interactions using a Markov chain, an idea that we adapt to capture the fact that users tend to browse art with consistent latent attributes during the course of a browsing session. VBPR is a recently-proposed content-aware recommender system, that models the visual appearance of the items being considered (primarily to recommend clothing). By combining the two models, we can capture individual users' preferences towards particular visual art styles, as well as the tendency to interact with items that are `visually consistent' during a browsing session. We also make several extensions of these models, e.g.~to handle longer memory than simply the previous action.

Another aspect that differentiates our model and data from traditional recommendation scenarios is the need to model the content \emph{creator} in addition to the content itself. Unlike most recommender systems, which model interactions between users and items, in \website{} the items are created by the same population of users who evaluate them. Thus we need to model the social dynamics that guide users' preferences toward particular artists. For example, a user might follow the creations of
particular artists, 
even when their art differs from the style they normally exhibit preferences toward.
Note that this particular type of social dynamics (amongst creators and evaluators) is different from traditional friendship/trust relations (amongst evaluators themselves).

We perform experiments on both implicit (click) and explicit 
(appreciate)
data, where we show that substantial performance improvements can be gained by modeling temporal, social, and visual features in concert. Improvements are especially large for newly-uploaded art, where modeling content---or even the identity of the artist---yields improvements of up to 30\% in terms of AUC.

Our model, 
\modelnamelong, or \md{} for short,
is the first to be simultaneously visually, socially, and temporally aware, which proves key to modeling the dynamics of online art communities.

\xhdr{Contributions} We summarize our main contributions as follows:
\begin{itemize}
\item We build models to capture the large-scale dynamics of artistic preferences. Methodologically we extend recent techniques that model visual appearance for content-aware recommendation, and techniques that model short-term, session-level dynamics, to capture the notion of `visual consistency' between successive actions. Since users of \website{} are themselves content creators, we capture both preferences toward particular artists, as well as particular art styles.
\item We introduce the \website{} dataset as a benchmark for modeling visual and artistic preferences. We perform experiments on both click and `appreciate' data, and while the former is proprietary, we shall release complete `appreciate' data for benchmarking and evaluation.
\item Our experiments reveal that each of the components (social, temporal, visual) are critical to obtaining satisfactory performance on our data, substantially outperforming models that consider each of the factors in isolation. We also use our model to visualize the various art styles that users exhibit preferences toward.
\end{itemize}
Note in particular that our proposed method can be easily adapted to cope with sequential prediction in other domains by replacing the visual features with other types of item content.

\section{Related Work}

\xhdr{Visually-aware recommender systems.}
In recent years, there have been some works that investigate parsing or retrieving visually similar clothing images (e.g.~\cite{yamaguchi2013paper,yang2014clothing,kalantidis2013getting}), although they do not focus on learning user preferences.
Recent works have introduced visually-aware recommender systems where the visual rating dimensions of users are uncovered on top of the visual signals in the system---product images. Such visual dimensions have been demonstrated to be successful at link prediction tasks useful for recommending alternative (e.g.~two similar t-shirts) and complementary (e.g.~a t-shirt and a matching pair of pants) items. This thread of work has also extended standard Matrix Factorization with visual dimensions to facilitate item recommendation tasks \cite{VBPR,sherlock}. A following work \cite{upsdowns} further considers the long-term temporal dynamics associated with the visual preference of the communities, i.e., fashion evolution. However, none of above works focuses on predicting session-level user actions, a key recommendation task that requires the ability to model \emph{sequential} data.

\xhdr{Modeling Temporal Dynamics.}
Apart from the visual domain, in the machine learning community there have also been earlier efforts that investigate temporally-evolving data with algorithms such as SVMs~\cite{klinkenberg2004learning}, decision trees~\cite{wang2003mining}, instance-based learning~\cite{aha1991instance}, etc. Similar to ours are Collaborative Filtering models that take into account temporal dynamics. In particular, early works are based on similarity-oriented Collaborative Filtering, where a time weighting scheme can be used to assign previously-rated items with decaying weights when computing similarities (e.g.~\cite{timeweightCF}). In contrast, recent works are mostly based on the well-known Matrix Factorization technique \cite{koren2008factorization}. For instance, Koren~\cite{timeSVD} showed state-of-the-art results on \emph{Netflix} data using Matrix Factorization to model the underlying temporal dynamics. However, these works do not consider handling the complex semantics of visual and social signals. 

\xhdr{Sequential recommendation.} 
Markov chains are a powerful tool to model stochastic transitions between different `states.' In sequential recommendation domains, Markov chains have been studied by several earlier works, from investigating their strength at uncovering sequential patterns (e.g.~\cite{zimdars2001using,mobasher2002using}), to directly modeling decision processes with Markov transitions \cite{shani2002mdp}.
In a more recent paper \cite{rendle2010fpmc}, Rendle \emph{et al.}~proposed to combine the strength of Markov chains at modeling the smoothness of subsequent actions and the power of Matrix Factorization at modeling \emph{personal} preferences for sequential recommendation. The resulting model, called FPMC, has shown superior prediction accuracy by benefiting from both simultaneously. 
Our work extends the basic idea mainly by modeling complex visual and social dynamics and further considering Markov chains with higher orders.

Notably, there is 
another line of work that employs (personalized) probabilistic Markov embeddings to model sequential data like playlists and POIs (e.g.~\cite{chen2012playlist,chen2013multi,feng2015personalized}). These works differ 
from ours in terms of both the signals being modeled and the types of models. Moreover, none of them has shown to be able to handle datasets on the same scale with \website{} to the best of our knowledge.

\xhdr{Social recommendation.} 
In the recommender systems literature, there has been a large body of work that models social networks for mitigating cold-start issues in recommender systems, e.g.~\cite{chaney2015probabilistic,zhao2014leveraging,ma2009learning,ma2008sorec,guo2015trustsvd}. The type of social signals they usually benefit from are so-called `trust' relations amongst different users, which are 
different from the those we are interested in.
In contrast, we focus on modeling the \emph{ownership} signal, i.e., the interactions between a viewer and the creator of an item, which brings a unique set of challenges.

\section{The Markov Chain Model} \label{sec:model}

\begin{table}
\centering
\renewcommand{\tabcolsep}{4pt}
\caption{Notation \label{tab:notation}}
\begin{tabular}{lp{0.75\linewidth}} \toprule

Notation & Explanation\\ \midrule
$\mathcal{U}, \mathcal{I}$ & user set, item set\\
$u, i$ & a specific user or item (resp.)\\
$\mathcal{S}^u$ & sequential action history $(\mathcal{S}^u_1, \ldots, \mathcal{S}^u_{|\mathcal{S}^u|})$ of $u$ \\
$\mathcal{S}^u_t$ & the item user $u$ interacted with at time step $t$\\
$o_i$ & owner/creator of item $i$, $o_i \in \mathcal{U}$ \\
$O_i$ & owners/creators of item $i$, $O_i \subseteq \mathcal{U}$ \\
$\Gamma$ & latent space measuring \emph{user-item} affinity \\
$\gamma_u$ & latent preference vector of user $u$, $\gamma_u \in \Gamma$ \\
$\gamma_i$ & latent property vector of item $i$, $\gamma_i \in \Gamma$ \\
$\Phi$ & latent space measuring \emph{user-user} similarity \\
$\phi_u$ & latent vector of user $u$, $\phi_u \in \Phi$ \\
$\Psi$ & latent space measuring \emph{item-item} similarity \\
$\psi_i$ & latent vector of item $i$, $\psi_i \in \Psi$ \\
$f_i$ & explicit feature vector of item $i$ (e.g.~visual features), $f_i \in \mathbb{R}^F$ \\
\bottomrule
\end{tabular}
\renewcommand{\tabcolsep}{6pt}
\end{table}

\subsection{Problem Formulation}
Formally, let $\mathcal{U} = \{u_1, u_2, \ldots, u_{|\mathcal{U}|}\}$ represent the set of users and $\mathcal{I} = \{i_1, i_2, \ldots, i_{|\mathcal{I}|}\}$ the set of items (i.e., artistic images).
Each item $i \in \mathcal{I}$ is created by a certain user $o_i \in \mathcal{U}$ (or in some cases multiple users $O_i \subseteq \mathcal{U}$). 
For each user $u$, a sequential action history (e.g.~items clicked/appreciated by $u$) $\mathcal{S}_u$ is known: $\mathcal{S}^u = (\mathcal{S}^u_1, \mathcal{S}^u_2, \ldots, \mathcal{S}^u_{|\mathcal{S}^u|})$ where $\mathcal{S}^u_k \in \mathcal{I}$. The action history of all users is denoted by $\mathcal{S} = \{\mathcal{S}^{u_1}, \mathcal{S}^{u_2}, \ldots, \mathcal{S}^{u_{|\mathcal{U}|}}\}$. Additionally, each item $i$ is associated with an explicit feature vector $f_i$, e.g.~in our case visual features extracted from images. Given the above data, our task is to recommend items to each user based on their predicted next action.

Notation to be used throughout the paper is summarized in Table \ref{tab:notation}. Next, we will build our model incrementally which we detail in each of the following subsections.

\subsection{Predicting Sequential Actions} \label{sec:basicmodel}
In this paper, we are interested in modeling \emph{binary} actions such as clicks, appreciates, etc. Given the previous action sequence $\mathcal{S}^u$ of user $u$, the probability that user $u$ will interact with item $i$ in the next action is $P(X^u_{t} = i~|~\mathcal{S}^u_{t-1}, \mathcal{S}^u_{t-2}, \ldots , \mathcal{S}^u_1)$, where random variable $X^u_{t}$ is the choice $u$ made at time step $t$ and $|\mathcal{S}^u| = t-1$. This probability can be approximated by $P(X^u_{t} = i | \mathcal{S}^u_{t-1})$, which is then modeled by factorizing a personalized Markov chain (as in \cite{rendle2010fpmc}). This follows the intuition that a user's \emph{recent} actions should be more predictive of their future actions than earlier ones.

Intuitively, the transition of user $u$ from item $\mathcal{S}^u_{t-1}$ (at time step $t-1$) to item $i$ (at time step $t$) can be explained from two aspects: (1) the interaction between user $u$ and item $i$, which captures the \emph{long-term} preferences of user $u$; and (2) the interaction between the previous item $\mathcal{S}^u_{t-1}$ and item $i$, which captures the \emph{short-term} or temporary interest of user $u$. Following this intuition, we propose to factorize the personalized Markov chain with the following formulation:
\begin{multline} \label{eq:basic}
 P(X^u_{t} = i~|~\mathcal{S}^u_{t-1}) \propto  
 \underbrace{\overbrace{\langle \phi_u, \phi_{o_i} \rangle}^{\substack{\text{owner}\\\text{appreciation}}} + \overbrace{\langle \gamma_u, \gamma_i \rangle}^{\substack{\text{item}\\\text{appreciation}}}}_{\text{interaction between $u$ and $i$ (\emph{long}-term dynamics)}}  \\
+ w_u \cdot (\underbrace{\overbrace{\langle \phi_{o_i}, \phi_{o_{\mathcal{S}^u_{t-1}}} \rangle}^{\text{owner similarity}} 
+ \overbrace{\langle \psi_i, \psi_{\mathcal{S}^u_{t-1}} \rangle}^{\text{item similarity}}}_{\mathclap{\text{interaction between $i$ and $\mathcal{S}^u_{t}$ (\emph{short}-term dynamics)}}}),
\end{multline}
where we use three latent spaces to capture the interactions between users and items, users and users, items and items respectively: 
(1) \textbf{Space $\Gamma$:} $D_1$-dimensional vectors $\gamma_u$ and $\gamma_i$ are employed to capture user $u$'s latent preferences and item $i$'s latent properties respectively;
(2) \textbf{Space $\Phi$:} Each user $u$ is associated with a $D_2$-dimensional vector $\phi_u$ for measuring affinity/similarity with other users (within $\Phi$);
and (3) \textbf{Space $\Psi$:} Each item $i$ is associated with a $D_3$-dimensional vector $\psi_i$ for measuring affinity/similarity with other items (within $\Psi$). 
Inner products of vectors from $\Gamma$, $\Phi$, and $\Psi$ measure the corresponding user-item, user-user, and item-item affinity/similarity respectively.

In \eq{eq:basic}, the interaction between $u$ and $i$ consists of two parts: how much user $u$ \emph{appreciates}---or is `similar' to---the creator/owner $o_i$ of $i$ (term one), and how much $u$ likes the \emph{specific} item created by $o_i$ (term two). Likewise, the similarity of $i$ and $\mathcal{S}^u_{t-1}$ comprises two components as well: the similarity between their creators (term three) and the two items themselves (term four). $w_u$ is a parameter to capture the statistical short-term \emph{consistency} of the actions of the specific user $u$. Intuitively, a bigger value of $w_u$ will favor short-term interests, which means $u$ tends to interact with items similar to those they just interacted with, instead of those that most fit their long-term preferences.

Modeling ownership data and user interactions can help address both \emph{cold user} and \emph{cold item} issues in real-world recommender systems: 
\begin{enumerate}
\item For a cold (or `cool') user $u$ who has very few actions in the system, the representation of $u$ (in space $\Phi$, i.e., $\phi_u$) can still be learned as long as some users have interacted with items \emph{created by} \textbf{$u$}. When predicting the actions for $u$, $\phi_u$ will help rank items created by similar users higher.
\item For a cold item $i$ with very few interactions in the system, we can still model it reasonably well as long as its creator's representation $\phi_{o_i}$ can be learned. Fortunately, two sources of signals are available in the system for inferring $\phi_{o_i}$: both actions made by $o_i$ (active) and the interactions received by all other items created by $o_i$ (passive). These signals are relatively abundant and unlikely to suffer from sparsity issues.
\end{enumerate}

\xhdr{Handling Multiple Creators.} In certain cases an item $i$ may be collaboratively created by multiple users in the system $O_i \subseteq \mathcal{U}$. In such scenarios we take the \emph{average} of their associated vectors as the corresponding vectors (i.e., $\phi_{o_i}$ and $\phi_{o_{\mathcal{S}^u_{t-1}}}$) in \eq{eq:basic}. Note that this will not affect the training efficiency as an owner sampling scheme can be used, which we will detail later.

\subsection{Modeling Higher-Order Markov Chains}
The formulation in \eq{eq:basic} only models a personalized Markov chain of order 1, with the assumption that a user's next action (at $t$) is independent of any historical actions if given the most recent one (at $t-1$). However, this suffers from noise and can't capture longer-term consistency (e.g.~earlier clicks in a session). This inspires us to address this limitation by modeling Markov chains with higher order.

Due to the large scale and sparsity of real-world data, a light-weight yet expressive model is required. To this end, we model high order personalized Markov chains by extending our first-order formulation with a personalized \emph{decaying} scheme, as shown below. The first part still measures the affinity between a user $u$ and the next item $i$ with long-term memory; while the second part computes the weighted sum of the similarities of item $i$ to each of the previous items:
\begin{equation} \label{eq:highorder}
\begin{split}
 P(X^u_{t} = i~|~\mathcal{S}^u_{t-1}, \mathcal{S}^u_{t-2},\ldots, \mathcal{S}^u_{t-K}) \propto  
 \underbrace{\langle \phi_u, \phi_{o_i} \rangle + \langle \gamma_u, \gamma_i \rangle}_{\text{interaction between $u$ and $i$}}
 \\ +
\sum^{K}_{k=1} \underbrace{w^k_u}_{\substack{\text{personalized}\\\text{decaying weight}}} \cdot~
(\underbrace{\langle \phi_{o_i}, \phi_{o_{\mathcal{S}^u_{t-k}}} \rangle + \langle \psi_i, \psi_{\mathcal{S}^u_{t-k}} \rangle}_{\text{interaction between $i$ and the $k$-th previous item}}).
\end{split}
\end{equation}

There are two main intuitions behind the proposed formulation: (1) recent actions should be more correlated with future actions, which is why we employ a decaying term; and (2) different users may differ in behavior so that personalization should be taken into account. We model the personalized decay scheme as follows:
\begin{equation}
w^k_u = \frac{1}{e^{a_u \cdot (k-1) + b_u}}, k = 1,2, \ldots ,K, \text{for}{\ } u \in \mathcal{U}. 
\end{equation}
Note that the above formulation only introduces two additional parameters (i.e., $a_u$ and $b_u$) for each user $u$, which allows us to model users' differing behavior in a light-weight manner.

\subsection{Incorporating Content-Based Features}
Up to now, our formulation only makes use of the collaborative data, without being aware of the underlying content of the items themselves.\footnote{Without loss of generality, we take item features as an illustrative example. User features can be handled in a similar manner.} Such a formulation may suffer from \emph{cold item} issues where there aren't enough historical observations to learn accurate representations of each item. Modeling the content of the items can provide auxiliary signals in cold-start settings and alleviate such issues.

Intuitively, content-based features of items should be informative of the two latent vectors: $\gamma_i$ and $\psi_i$. Given the explicit feature vector $f_i$ of item $i$, embedding techniques have been successful at incorporating (high dimensional) content-based features into the Collaborative Filtering framework (e.g.~\cite{VBPR,upsdowns}).
In particular, we augment the vector representations of items (i.e., $\gamma_i$, $\psi_i$) as follows:
\begin{eqnarray}
\gamma_i &= \gamma'_i + \mathbf{E}_\Gamma f_i, \label{eq:extendgamma}\\
\psi_i &= \psi'_i + \mathbf{E}_\Psi f_i. \label{eq:extendpsi}
\end{eqnarray}
Here $\mathbf{E}_\Gamma$ and $\mathbf{E}_\Psi$ are two embedding matrices that project item $i$ into space $\Gamma$ and $\Psi$ respectively. The new formulation for each vector in the corresponding space ($\gamma_i$) now consists of two parts: the base part from the item content ($\mathbf{E}_\Gamma f_i$), and the residue ($\gamma'_i$). For warm-start items, the residue part is expressive and can represent the item accurately; for cold-start items, the residue part will be regularized (towards 0) and the base part will still be able to provide reasonably good approximations of the true representations. By substituting the augmented vector representations \eq{eq:extendgamma} and \eq{eq:extendpsi} into \eq{eq:highorder}, the new formulation will be able to benefit from the content-based features to alleviate \emph{cold item} issues.

Note that the two matrices are shared by all items which means that only a modest number of parameters are added to the model.

\subsection{Fitting the Model}
Predicting next actions can be viewed as a ranking task where the ground-truth item should be ranked as high as possible among all items. This inspires us to optimize an efficient ranking loss such as sequential Bayesian Personalized Ranking (S-BPR) \cite{rendle2010fpmc}.\footnote{S-BPR was originally proposed to optimize sequential \emph{set} data. However, the same idea can also be applied to single-item cases.}

\subsubsection{Objective Function}
Let $>_{u,t}$ denote the total order over the items of user $u$ at time step $t$, then $i >_{u,t} j$ means item $i$ is ranked higher than item $j$ for user $u$ at $t$ given the action sequence before $t$. S-BPR optimizes \emph{maximum a posteriori} (MAP) of all the model parameters (denoted by $\Theta$) assuming independence of users and time steps:
\begin{equation}
\argmax_{\Theta} = \ln \prod_{u \in \mathcal{U}} \prod_{t = 2}^{|\mathcal{S}^u|} \prod_{j \neq \mathcal{S}^u_t} P(\mathcal{S}^u_t >_{u,t} j | \Theta) ~ P(\Theta),
\end{equation}
where for each user $u$ and for each time step $t$, the objective goes through the \emph{pairwise} ranking between the ground-truth $\mathcal{S}^u_t$ and all negative items. The pairwise ranking between a positive ($i$) and a negative ($j$) item $P(i >_{u,t} j | \Theta)$ is estimated by a sigmoid function $\sigma(\widehat{p}_{u,t,i} - \widehat{p}_{u,t,j})$ where the difference of the predicted likelihoods are taken as the inputs ($\widehat{p}_{u,t,i}$ is short for the Markov prediction, i.e., \eq{eq:highorder}). $P(\Theta)$ is a Gaussian prior over the model parameters.

Note that for Markov chains of order $K$, our formulation (i.e., \eq{eq:highorder}) can allow $t$ to run from $2$ (instead of $K+1$) to the last item in $\mathcal{S}^u$ due to its additive characteristic. 

\subsubsection{Scaling Up}
Stochastic Gradient Descent (SGD) is widely employed for learning in BPR-like optimization frameworks (e.g., \cite{rendle2009bpr,rendle2010fpmc}). According to SGD, our basic training procedure is as follows. First, a user $u \in \mathcal{U}$ as well as a time step $t \in \{2,3,\ldots,|\mathcal{S}^u|\}$ are uniformly sampled from the training corpus. Next, a negative item $j \in \mathcal{I}$ ($j \neq \mathcal{S}^u_t$) is uniformly sampled, which forms a training triple $(u,t,j)$. Finally, the optimization procedure updates parameters as follows: 
\begin{equation} \label{eq:sgd}
\Theta \leftarrow \Theta + \alpha \cdot (\sigma(-\widehat{x}_{u,t,j}) \cdot \frac{\partial \widehat{x}_{u,t,j}}{\partial \Theta} - \lambda_{\Theta}\Theta ), 
\end{equation}
where $\widehat{x}_{u,t,j}$ is a shorthand for $\widehat{p}_{u,t,\mathcal{S}^u_t} - \widehat{p}_{u,t,j}$. $\alpha$ is the learning rate and $\lambda_{\Theta}$ is a regularization hyperparameter. Note that it is cheap to evaluate the gradient due to the additive characteristic of the proposed method (see Appendix for complexity analysis).

The above procedure is suitable for training on large datasets. However, it is limited when training with the augmented version of our model (with content-based features) as updating the embedding matrices can be time-consuming. To speed up the training procedure, we make the following two observations and employ two modifications accordingly.

\xhdr{Sampling.} Matrices $\mathbf{E}_\Gamma$ and $\mathbf{E}_\Psi$ are \emph{global} parameters and only account for a tiny fraction of $\Theta$ (e.g.~$0.29\%$ for \website). 
This means that less training data is needed to accurately estimate $\mathbf{E}_\Gamma$ and $\mathbf{E}_\Psi$. In other words, they are updated more often than needed under the above scheme.
As such, we lower their updating frequency by flipping a biased coin for each training triple to decide whether or not to update $\mathbf{E}_\Gamma$ and $\mathbf{E}_\Psi$. Likewise, we can also sample a single owner and only update their associated parameters in multiple-owner cases due to the rich user interactions available.

\xhdr{Asynchronous SGD.} Notably, only a tiny fraction of parameters will be updated for each training triple $(u,t,j)$, i.e., representations of $u$ and a few \emph{relevant} items, and they are unlikely to overlap.\footnote{The sampling scheme also helps reduce collisions when updating the embedding matrices.} It has been pointed out that in such cases lock-free parallelization of SGD could be employed to achieve fast convergence \cite{recht2011hogwild}. 

Experimentally, this na\"{\i}ve sampling and asynchronous SGD procedure can help finish training on huge datasets within reasonable time on commodity machines without losing prediction accuracy.

\section{Experiments}
We perform comprehensive experiments on \website{} to evaluate the proposed method. We begin by introducing the dataset we work with, and then we compare against a variety of baselines. Finally, we visualize the learned model and qualitatively analyze the results. 

\subsection{Dataset}
\website{} is a popular online community where millions of professional photographers, designers and artists share their work, organized as projects, with others. The contents of these projects vary significantly, ranging from photographs to animation, fashion, interior design, paintings, sculpting, calligraphy, cartoons, culinary arts, etc. Each project is created by a user or a few users and consists of a collection of images (as well as videos in certain cases). The creator/owner of the project selects the most representative image which the website presents to all users as the cover image. On the website, users browse through large numbers of cover images, click through attractive projects, and `appreciate' those they like.

From \website{} we collected two large corpora of timestamped user actions: (1) \emph{appreciates} consisting of 373,771 users, 982,002 items (i.e., projects), and 11,807,103 appreciates (i.e., explicit feedback), and (2) \emph{clicks} comprising 381,376 users, 972,181 items, and 48,118,748 clicks (i.e., implicit feedback). 52.7\% users have created their own projects, and 2.3\% items are created by multiple users. In our experiments, appreciates and clicks are used as two separate datasets to evaluate the efficacy of all methods on explicit and implicit datasets respectively. 

For each item, we extract a 4096-dimensional feature vector from its cover image with a pre-trained VGG neural network \cite{simonyan2014vgg} as the content-based features $f_i$. Such features have been shown to generate state-of-the-art results on visually-aware item recommendation tasks \cite{VBPR,upsdowns}.

Complete appreciate data and code are available on the first author's webpage.\footnote{\url{https://sites.google.com/a/eng.ucsd.edu/ruining-he/}}

\subsection{Evaluation Metric} \label{sec:metric}
We construct a validation set and a test set by selecting the most recent two actions of each user, one for validation and the other for testing. The remainder are used for training. 
All comparison methods were trained on the training set with hyperparameters tuned with the validation set. Finally, the trained models are used to report corresponding performance on the test set.

Recall that the task is to predict which item a user will interact with given the previous action history, which means that the model in question needs to rank the ground-truth item higher than other items. Therefore for the held-out action, a reasonable evaluation would be calculating how highly the ground-truth item has been ranked for each user. 

To this end, we evaluate all methods on the \emph{test} set with the widely used AUC (\emph{Area Under the ROC curve}) measure (e.g., \cite{VBPR,upsdowns,rendle2009bpr}):
\begin{equation}
\mathit{AUC} =  \frac{1}{|\mathcal{U}|}  \sum_u   \frac{1}{|\mathcal{I}\setminus\mathcal{I}^+_u|}   \sum_{j \in \mathcal{I} \setminus \mathcal{I}^+_u}  \mathbf{1} (\widehat{p}_{u,|\mathcal{S}^u|,g_u} > \widehat{p}_{u,|\mathcal{S}^u|,j}),
\end{equation}
where $\mathbf{1}(\cdot)$ is the indicator function,
and $g_u$ is the ground-truth item of user $u$ at the most recent time step $|\mathcal{S}^u|$.

\subsection{Comparison Methods}
We compare against a series of state-of-the-art methods, both in the field of item recommendation and sequential data prediction.

\xhdr{Popularity (POP):} always recommends popular items in the system at each time step.

\xhdr{Bayesian Personalized Ranking (BPR-MF) \cite{rendle2009bpr}:} is a state-of-the-art method for personalized item recommendation. It takes standard Matrix Factorization \cite{korenSurvey} as the underlying predictor.

\xhdr{Visual Bayesian Personalized ranking (VBPR) \cite{VBPR}:} is an extension of BPR-MF that models raw visual signals for item recommendation. It captures the visual but not the temporal dynamics of binary action sequences.

\xhdr{Factorized Markov Chain (MC):} factorizes the $|\mathcal{I}| \times |\mathcal{I}|$ transition matrix to capture the likelihood that a user transits from one item to another. We used a first order model as higher orders incur a state-space explosion (we have almost one million items) and degrade the performance, especially considering the data sparsity.

\xhdr{Factorized Personalized Markov Chain (FPMC) \cite{rendle2010fpmc}:} is a state-of-the-art personalized Markov chain algorithm to model sequential data. However, FPMC is unable to capture visual and social dynamics and only models first-order Markov chains.

\xhdr{\modelnamelong{} (\md):} is the method proposed by this paper in \eq{eq:highorder}. Markov chains of different orders will be experimented with and compared against other methods. 

\xhdr{\md+:} augments \md{} with the 4096-dimensional visual features extracted from cover images.

\begin{table}
\begin{center}
\renewcommand{\tabcolsep}{2pt}
\caption{Models}\label{table:base}
\begin{tabular}{lccccc} \toprule
Model       &\parbox{0.18\linewidth}{\centering Personal-ized?} &
\parbox{0.18\linewidth}{\centering Temporally-aware?} &
\parbox{0.18\linewidth}{\centering Socially-aware?} & \parbox{0.18\linewidth}{\centering Visually-aware?}   \\ \midrule 
POP         &\ding{56} &\ding{56} &\ding{56}   &\ding{56}   \\  
BPR-MF      &\ding{52} &\ding{56} &\ding{56}   &\ding{56}   \\
VBPR        &\ding{52} &\ding{56} &\ding{56}   &\ding{52} \\
MC          &\ding{56} &\ding{52} &\ding{56}   &\ding{56}   \\
FPMC        &\ding{52} &\ding{52} &\ding{56}   &\ding{56}   \\ 
\md         &\ding{52} &\ding{52} &\ding{52}   &\ding{56}   \\ 
\md+        &\ding{52} &\ding{52} &\ding{52}   &\ding{52} \\ \bottomrule
\end{tabular}
\end{center}
\end{table}

\begin{table*}
\centering
\renewcommand{\tabcolsep}{3.2pt}
\caption{AUC for next-\textbf{appreciate} prediction (higher is better). `Full' evaluates the overall accuracy, `\emph{Cold User}' measures the ability to recommend/rank items for cold users, and `\emph{Cold Item}' examines the ability to recommend cold items correctly. The rest are more detailed settings for evaluation. The best performance for each setting is boldfaced.
We test different orders of the Markov chain (i.e.,~1, 2, and 3).}
\begin{tabular}{lcccccccccccccccccccc} \toprule
\multirow{2}{*}{Setting} &(a)    &(b)    &(c)   &(d)    &(e)    &(f-1)  &(f-2)  &(f-3) &(g-1)   &(g-2)  &(g-3)  &\multicolumn{2}{c}{improvement}\\ 
                        &POP    &BPR-MF  &VBPR  &MC     &FPMC   &\md    &\md    &\md   &\md+    &\md+   &\md+   &ours vs.~e &ours vs. best\\ \midrule
Full                    &0.8282 &0.8559 &0.8802 &0.8449 &0.8689 &0.9199 &0.9220 &0.9206 &0.9203 &\textbf{0.9270} &0.9203 & 6.69\%  & 5.32\% \\
\emph{Cold User}        &0.7637 &0.8132 &0.8500 &0.8134 &0.8208 &0.8994 &\textbf{0.9008} &0.8979 &0.8964 &0.8877 &0.8958 & 9.75\%  & 5.98\% \\ 
\emph{Cold Item}        &0.0562 &0.4384 &0.5826 &0.4559 &0.5429 &0.7574 &0.7644 &0.7604 &0.7686 &\textbf{0.7714} &0.7603 & 42.09\% & 32.41\% \\ [2.5pt]
Owner Trans.            &0.8470 &0.8814 &0.8960 &0.8600 &0.8800 &0.9076 &0.9115 &0.9118 &0.9074 &\textbf{0.9186} &0.9115 & 4.39\%  & 2.52\% \\  
Same Owner              &0.7677 &0.7388 &0.8076 &0.7753 &0.8176 &0.9761 &0.9699 &0.9609 &\textbf{0.9795} &0.9657 &0.9607 & 19.80\% & 19.80\% \\ 
Session Trans.          &0.7876 &0.8476 &0.8710 &0.8289 &0.8524 &0.9032 &0.9057 &0.9051 &0.9025 &\textbf{0.9126} &0.9045 & 7.06\%  & 4.78\% \\ 
Same Session            &0.8545 &0.8674 &0.8931 &0.8672 &0.8919 &0.9431 &0.9445 &0.9421 &0.9451 &\textbf{0.9472} &0.9424 & 6.20\%  & 6.06\% \\ \bottomrule
\end{tabular}
\label{table:auc_appre}
\end{table*}


\begin{table*}
\centering
\renewcommand{\tabcolsep}{3.2pt}
\caption{AUC for next-\textbf{click} prediction (higher is better). The best performance for each setting is boldfaced.}
\begin{tabular}{lcccccccccccccccccccc} \toprule
\multirow{2}{*}{Setting} &(a)    &(b)    &(c)   &(d)    &(e)    &(f-1)  &(f-2)  &(f-3) &(g-1)   &(g-2)  &(g-3)  &\multicolumn{2}{c}{improvement}\\ 
                        &POP    &BPR-MF  &VBPR  &MC     &FPMC   &\md    &\md    &\md   &\md+    &\md+   &\md+   &ours vs.~e &ours vs. best\\ \midrule
                 
Full                    &0.8116 &0.8972 &0.8968 &0.8954 &0.9089 &0.9319 &\textbf{0.9362} &0.9349 &0.9330 &0.9337 &0.9354 & 3.00\%  & 3.00\% \\
\emph{Cold User}        &0.7555 &0.8475 &0.8535 &0.8708 &0.8605 &0.9114 &\textbf{0.9154} &0.9127 &0.9121 &0.9114 &0.9126 & 6.38\%  & 5.12\% \\ 
\emph{Cold Item}        &0.1919 &0.4284 &0.5097 &0.5732 &0.4828 &0.6932 &0.7234 &0.7290 &0.7001 &\textbf{0.7398} &0.7367 & 53.23\% & 27.6\% \\ [2.5pt]
Owner Trans.            &0.8522 &0.9021 &0.9009 &0.8733 &0.9018 &0.9194 &0.9223 &0.9231 &0.9193 &0.9185 &\textbf{0.9232} & 2.37\%  & 2.34\% \\  
Same Owner              &0.6251 &0.8810 &0.8834 &0.9662 &0.9316 &0.9763 &0.9771 &0.9738 &0.9770 &\textbf{0.9825} &0.9747 & 5.46\%  & 1.69\% \\ 
Session Trans.          &0.8031 &0.8651 &0.8671 &0.8279 &0.8641 &0.9007 &0.9012 &0.9012 &0.9009 &0.8962 &\textbf{0.9013} & 4.31\%  & 4.18\% \\ 
Same Session            &0.8234 &0.9179 &0.9160 &0.9391 &0.9379 &0.9537 &0.9573 &0.9579 &0.9538 &\textbf{0.9580} &0.9575 & 2.14\%  & 2.01\% \\ \bottomrule
\end{tabular}
\label{table:auc}
\end{table*}

For clarity, the above methods are collated in Table \ref{table:base} in terms of whether they are `personalized,' `temporally-aware,' `socially-aware,' and `visually-aware.'
The ultimate goal is to evaluate (1) the performance of state-of-the-art item recommendation methods on our task (i.e., BPR-MF); (2) the gains from using raw visual signals especially for cold-start settings (i.e., VBPR); (3) the importance of modeling temporal dynamics (i.e., MC); (4) 
the effect of modeling personalized temporal dynamics (i.e., FPMC); 
and (5) the importance of modeling social dynamics (i.e., \md) and further performance enhancements from incorporating content-based features (i.e., \md+).

We performed all experiments on a single machine with 8 cores and 64GB main memory. Our own most time-consuming method---\md+ with a third-order Markov chain---required around 50 hours of training time.

\subsection{Performance \& Analysis} \label{sec:performance}
As analyzed in Section \ref{sec:basicmodel}, the proposed model should be helpful for cold-start settings including both cold user
and cold item
scenarios. Therefore, we compare all methods in terms of overall accuracy (denoted by `Full') as well as the two cold settings. Overall accuracy is evaluated with the \emph{full} test set as introduced in Section \ref{sec:metric}. `\emph{Cold User}' is evaluated by a subset of the full test set, consisting of only those cold users with at most $5$ actions in the training set; likewise, `\emph{Cold Item}' uses the subset comprising only cold items with at most $5$ interactions. 

In addition, we also decompose the full test set into subsets according to whether the previous item (at $t-1$) of the action being tested (at $t$) is created by a different owner (i.e., owner transition), or whether the action is the start of a new session\footnote{Since no session metadata is available, sessions are obtained by temporally partitioning each user's clicks/appreciates with gaps larger than 1hr.} (i.e., session transition). This gives us four settings: `Owner Trans.'~vs.~`Same Owner,' and `Session Trans.'~vs.~`Same Session' to thoroughly evaluate the ability of all models under various transitioning circumstances. 

For all methods, we try to use the same number of dimensions: BPR-MF and MC use 20 latent dimensions; VBPR uses 10 latent plus 10 visual dimensions; FPMC uses two 10-d vectors to represent each item and one 10-d vector to represent each user; \md{} and \md+ use 10 dimensions for $\gamma_u$, $\phi_u$, $\gamma_i$, and $\psi_i$ (i.e., $D_1 = D_2 = D_3 = 10$). Using additional dimensions yielded only marginal performance improvement for all methods. 

AUC results on the two datasets are shown in Tables \ref{table:auc_appre} and \ref{table:auc}. We make a few comparisons and analyze our findings as follows:

\xhdr{BPR-MF vs.~MC vs.~FPMC.} 
Ultimately, BPR-MF and MC focus on modeling long-term and short-term dynamics respectively. BPR-MF ranks items according to what the given user likes from a long-term perspective, which makes it relatively strong when a user's action differs significantly from the previous one (`Owner Trans.'~and `Session Trans.'). In contrast, MC ranks items based on the transition matrix, i.e., `similarity' of the next item to the previous item. Such short-term awareness makes MC strong in cases where action consistency is maximally demonstrated, i.e., `Same Owner' and `Same Session.' Additionally, note that MC seems to suffer less from cold-start issues due to the consistency of sequential actions. 
FPMC is inherently a combination of BPR-MF and MC, which makes it the strongest among the three, though it is not necessarily the best in all settings.

\xhdr{BPR-MF vs.~VBPR.}
Like BPR-MF, VBPR focuses on long-term dynamics, but benefits from making use of additional visual signals in the system. As expected, it alleviates \emph{cold item} issues and is slightly better than BPR-MF in most other settings as well.

\xhdr{\md{} vs.~FPMC.}
\md{} is able to capture social dynamics by modeling ownership signals as well as gains from a fully personalized Markov chains with higher orders. As such, it beats FPMC in all settings significantly especially in cold-start scenarios. Note that \md{} improves as much as 47.66\% on average for cold item recommendation, which is a major concern when predicting sequential actions (see Section \ref{sec:basicmodel} for detailed cold-start analysis of \md{}).

\xhdr{\md+ vs.~\md{}.}
Further augmenting \md{} with content-based features, in this case visual signals, is beneficial as \emph{cold item} issues are further mitigated. However, the improvement is comparatively small as such issues have already been alleviated to a large extent by modeling social dynamics (i.e., \md).

\begin{figure}
\centering
\includegraphics[width=\linewidth]{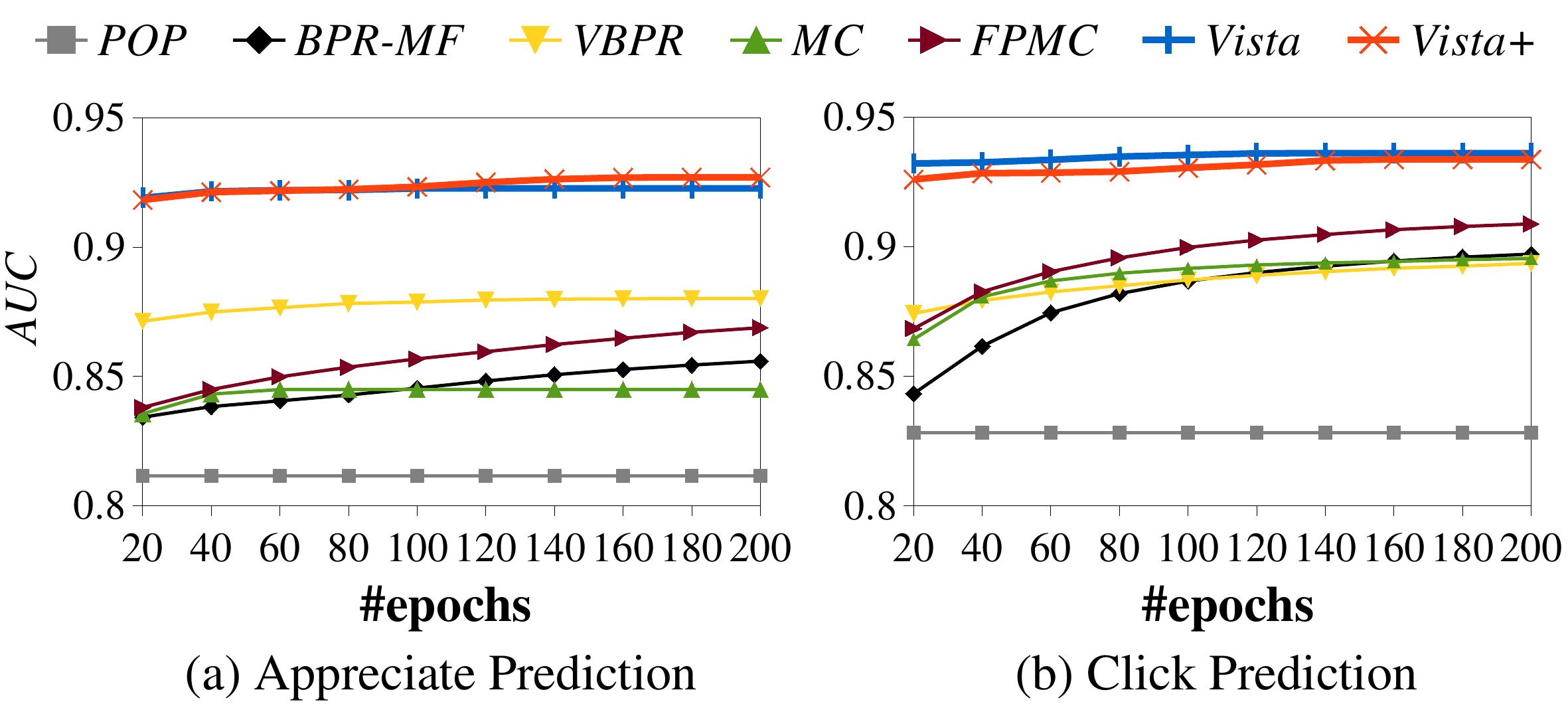}
\caption{Performance comparison of different methods with number of training epochs on \website{} data. All methods are from Section \ref{sec:performance}. \md{} and \md+ use second-order Markov chains.}
\label{fig:curve}
\end{figure}

\begin{figure*}[!t]
\centering
\includegraphics[width=\linewidth]{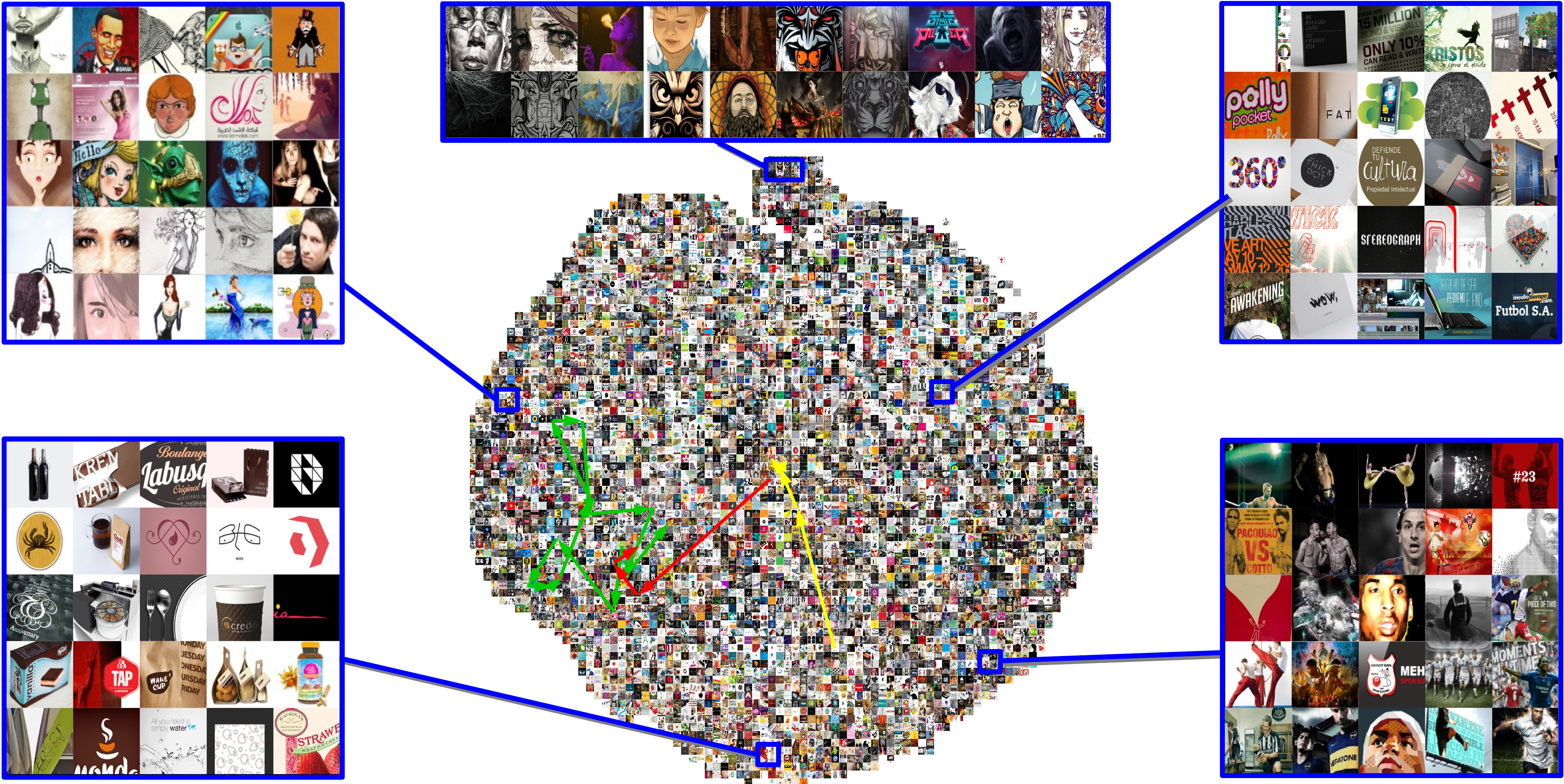}
\caption{A 2-d visualization of space $\Psi$ with t-SNE \cite{tsne}. Here the \md+ model is trained on \website{} click data with the same setting as in Table \ref{table:auc}. We randomly sample a user and visualize three of their click sessions in the space, denoted by three paths with different colors on the grid view. There is a tendency that clicks are around a specific region in the space (i.e., near the green arrows), which reflects the long-term preference of the user. And each click sequence shows a certain amount of consistency, encoding the transition of short-term interests.}
\label{fig:tsne}
\end{figure*}

\xhdr{Orders of Markov chains.}
As we increase the order of the Markov chain (from 1 to 3), the performance generally gets better as the next click is related to multiple previous clicks. Using a small order seems to be good enough, presumably since the few most recent actions capture enough information to predict future actions.

\xhdr{Sparse vs.~Dense data.} 
Comparing the results of next-appreciate (sparse) and next-click (dense) prediction we can find that \md{} outperforms other methods more significantly for \emph{sparse} datasets where social and visual dynamics are
forced to carry more weight.

\subsection{Training Efficiency}
Next, we demonstrate the accuracy on the full test set (i.e., `Full' setting) of all comparison methods as the number of training epochs increases.\footnote{All baselines, except POP, are trained with multiple epochs until convergence or observance of overfitting (on the validation set). Each epoch consists of processing training samples with the size of the whole training corpus.} As shown by Figure \ref{fig:curve}, our proposed methods can converge in a few epochs due to the rich interactions being modeled. Since \emph{appreciate} data is inherently sparser than \emph{click} data, we can find that modeling signals like social dynamics and visual data can help more significantly for appreciate prediction (as shown by the gap between (1) \md/\md+ vs.~others, and (2) VBPR vs.~other baselines). In our experiments, we used a sampling probability of 5\% for \md+ to update the two embedding matrices. Benefiting from multithreading, each epoch of asynchronous SGD takes around 20 minutes to train our third-order \md+ on click data. See Appendix for more details about the time complexity of \md+.

\subsection{Latent Space Visualization}
We proceed by visualizing the latent space $\Psi$, which is used to measure the similarity between different items. To this end, we take the \md+ trained on \website{} click data (see Section \ref{sec:performance}) and further use t-SNE \cite{tsne} to embed the 10-d space into 2-d. Figure \ref{fig:tsne} demonstrates a grid view of 972,181 items in the 2-d space.\footnote{Each cell randomly selects one image if items overlap.} From this figure we can see that items with similar contents and styles tend to be neighbors in the latent space, e.g.~culinary arts on the lower-left patch, and sports on the lower-right. Notably, items on the same patch may visually deviate from each other significantly. 

On the grid view, we also demonstrate a few click sessions of a randomly selected user. The click sequence of each session is represented by a directed path with a unique color (red/yellow/green).
We make a few observations as follows. (1) Clicks tend to occur around a specific region in the space, near the green and red arrows. This reflects the long-term preferences of the user as people ultimately tend to explore items that they like. (2) Each click sequence demonstrates consistency to a certain degree (e.g.~red and yellow) and encodes the transition of short-term interests (e.g.~green). (3) Finally, the choice made at each click is a combination of long- and short-term preferences, due to which there are both long jumps and short jumps. Therefore, it is essential to capture both long- and short-term dynamics simultaneously in order to be successful at addressing our prediction task. 

\subsection{Visualizing Sessions}
Users differ in habitual patterns especially in a dataset as large as ours. For example, for some users their short-term dynamics are more important. This means they tend to click/appreciate items similar to those they just interacted with. In contrast, there are also users whose long-term preferences are more emphasized, demonstrating less short-term consistency during a session. This characteristic is captured by the personalized weighting factor $w_u$ in \eq{eq:basic} (and $w^k_u$ in \eq{eq:highorder}). 

\begin{figure}[!t]
\centering
\includegraphics[width=\linewidth]{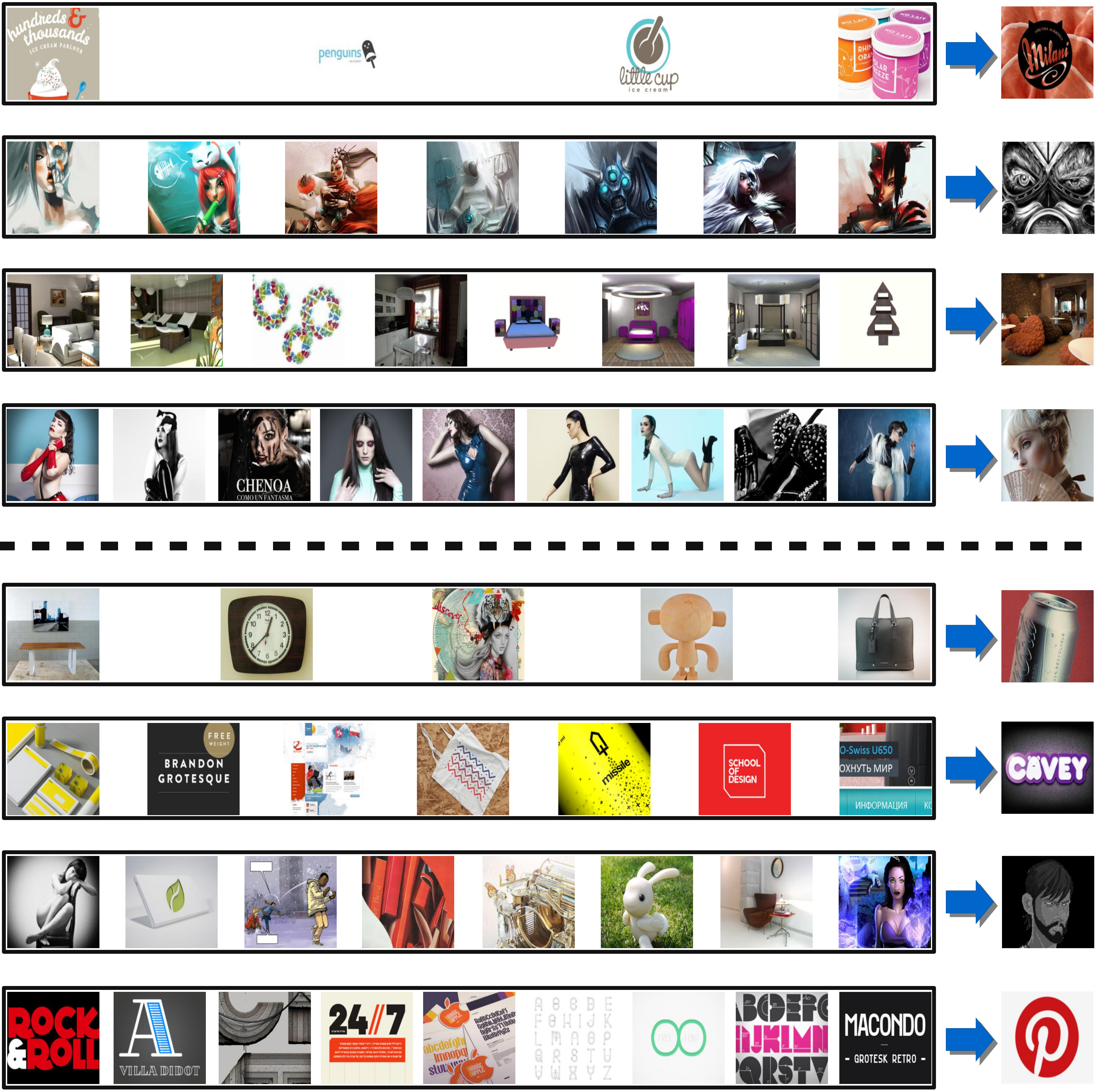}
\caption{Above the dashed line are four sample sessions of users with the largest $w_u$ (i.e., strong short-term consistency). Below the dashed line are four sample sessions of users with the least $w_u$ (i.e., weak short-term consistency). Items on the right of the arrows are recommendations made by \md+.}
\label{fig:session}
\end{figure}

In Figure \ref{fig:session}, we show a few sample sessions of the above two types of users, with different session lengths. Sampled sessions of users with the largest $w_u$ (i.e., $\argmax_u{w_u}$) are shown above the horizontal dashed line, with each row demonstrating the list of items clicked during the corresponding session. We can see some consistency from these sessions: logo designs, a certain style of cartoon characters, interior designs, and fashion models respectively. On the right of each arrow is the corresponding recommendation with the highest score predicted by the \md+ model. 

In contrast, below the dashed line we also demonstrate a few sessions from users with the least $w_u$, i.e., for whom long-term dynamics are more important. As expected, contents in each session demonstrate comparatively larger variance, though the long-term preference towards object designs, logos, cartoons, and characters (respectively) are captured by the \md+ model.

\section{Conclusion}
Modeling artistic preferences with complex visual, social, and sequential signals is challenging especially when it comes to the need to scale up to large real-world datasets. In this paper, we address these challenges by building visually and socially-aware Markov chains to model visual appearance and social dynamics simultaneously. Empirically we evaluated our proposed method on two common sequential prediction tasks to test its ability to handle both explicit and implicit artistic preferences of users. Experimental results demonstrated that our proposed methods significantly outperform a series of state-of-the-art baselines for both tasks on large scale datasets collected from a popular social art website, \website.

\appendix
\section{Scalability Analysis}
We analyze the time complexity of the $K$-order \md+ as follows. Recall that we use a sampling scheme to handle multiple creators (see Section \ref{sec:basicmodel}). For a given user $u$, time step $t$, and item $i$, the prediction $\widehat{p}_{u,t,i}$ takes $O(D_1*F+D_2)$ to compute the interactions between $u$ and $i$, and $O(K*(D_2+D_3*F))$ for the interactions between $i$ and $K$ previous items. Therefore for each sampled training triple $(u,t,j)$, the calculation of $\widehat{x}_{u,t,j}$ will take $O(D_1*F + D_2 + K*(D_2+D_3*F)) = O(D_1*F+K*D_2+K*D_3*F)$. According to \eq{eq:sgd}, computing the partial derivatives and updating relevant parameters also have the time complexity of $O(D_1*F + K*D_2+K*D_3*F)$ (when the two embedding matrices are updated). Summing up all above components, it takes $O(D_1*F+K*D_2+K*D_3*F)$ to process each training triple. Using our lock-free asynchronous SGD training procedure with sampling, multiple triples are trained simultaneously.

\setlength{\bibsep}{0pt}
\raggedright
\bibliographystyle{abbrv}
\bibliography{sigproc}

\end{document}